\documentclass[twocolumn,letterpaper]{revtex4}
\usepackage{graphicx,bm}
\usepackage[totalwidth=7in,totalheight=9.6in]{geometry}
\begin{document}
\title{Is small-world network disordered?}
\author{Soumen Roy}
\email{sroy@iopb.res.in}
\author{Somendra M. Bhattacharjee}
\email{somen@iopb.res.in}
\affiliation{Institute of Physics, Bhubaneswar 751005, India}
\def\Sa{Self-averaging }
\def\sa{self-averaging }
\def\bege{\begin{eqnarray}}
\def\ende{\end{eqnarray}}

\newcommand{\DIR}{./figs}
\newcommand{\BINDER}{%
\begin{figure}[htbp]
\centering
\includegraphics[width=2.0in]{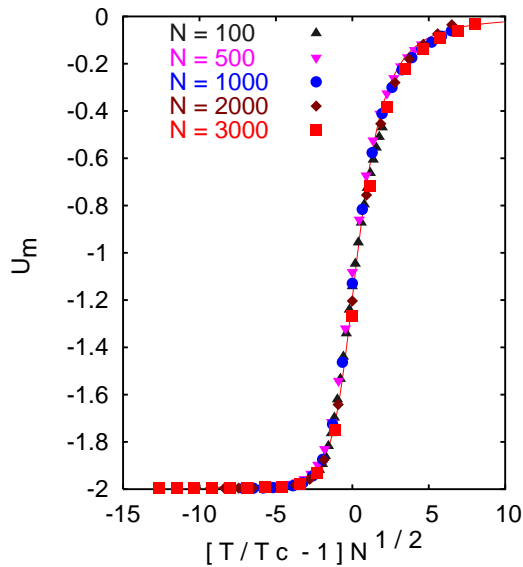}
\caption{Plot of the data-collapse of the Binder Cumulant versus
  scaled temperature for various $N$} 
\label{fig:BINDER}
 \end{figure}}

\newcommand{\DISTRIBUTION}{%
\begin{figure}[htbp]
  \centering
\includegraphics[width=2.0in]{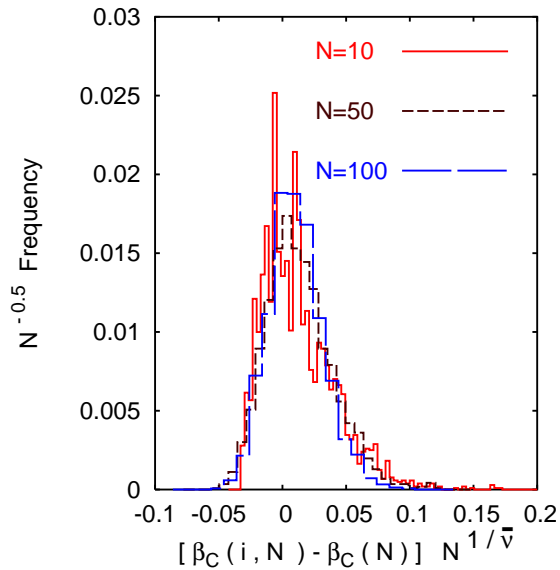}
\caption{Plot of the distribution of pseudocritical inverse
  temperatures $\beta_C(i ,N)$ for various $N$} 
\label{fig:DISTRIBUTION}
 \end{figure}}

\newcommand{\SELFAVG}{%
\begin{figure}
\begin{center}
\begin{tabular}{cc}
\resizebox{32mm}{!}{\includegraphics{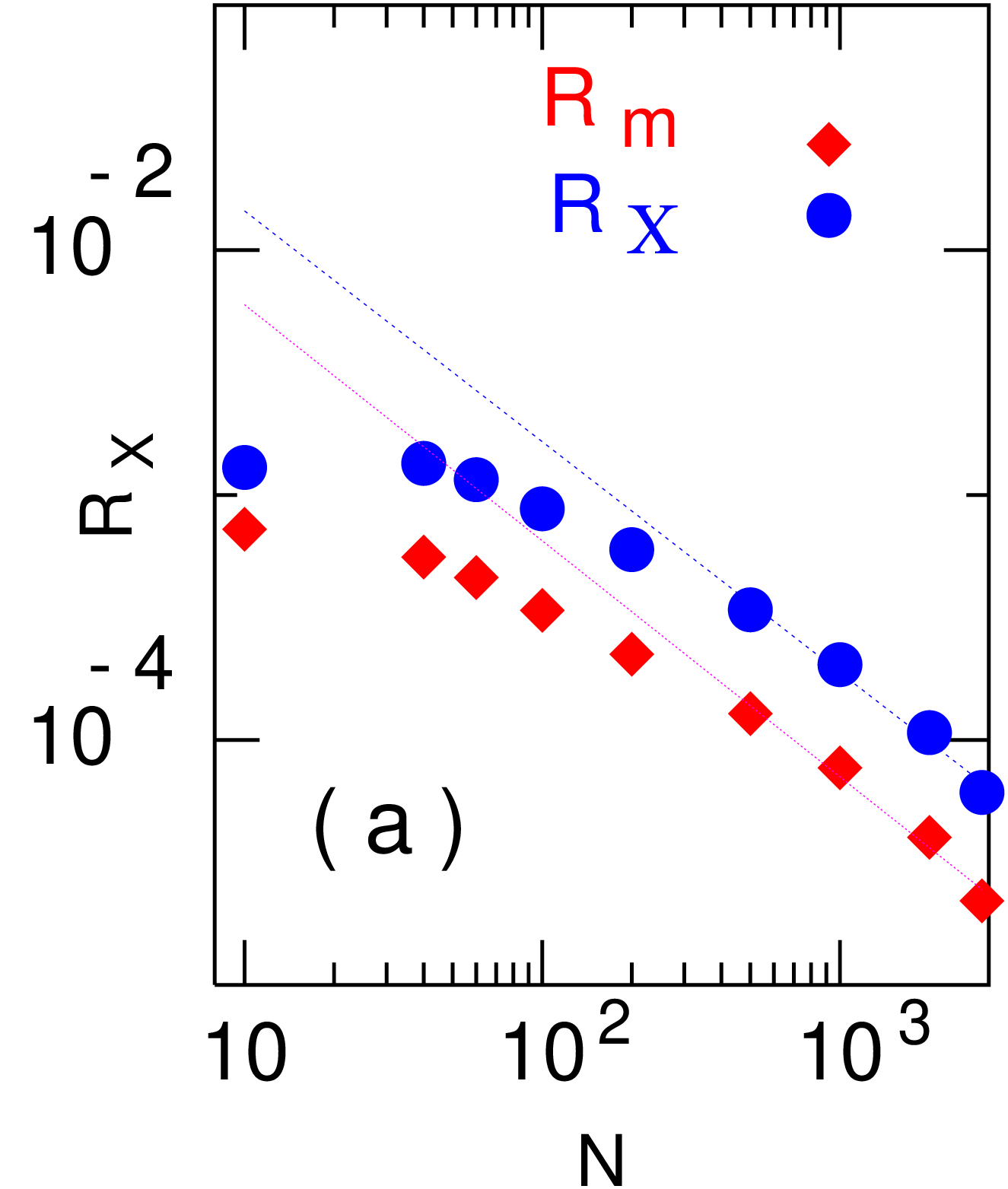}}
\resizebox{32mm}{!}{\includegraphics{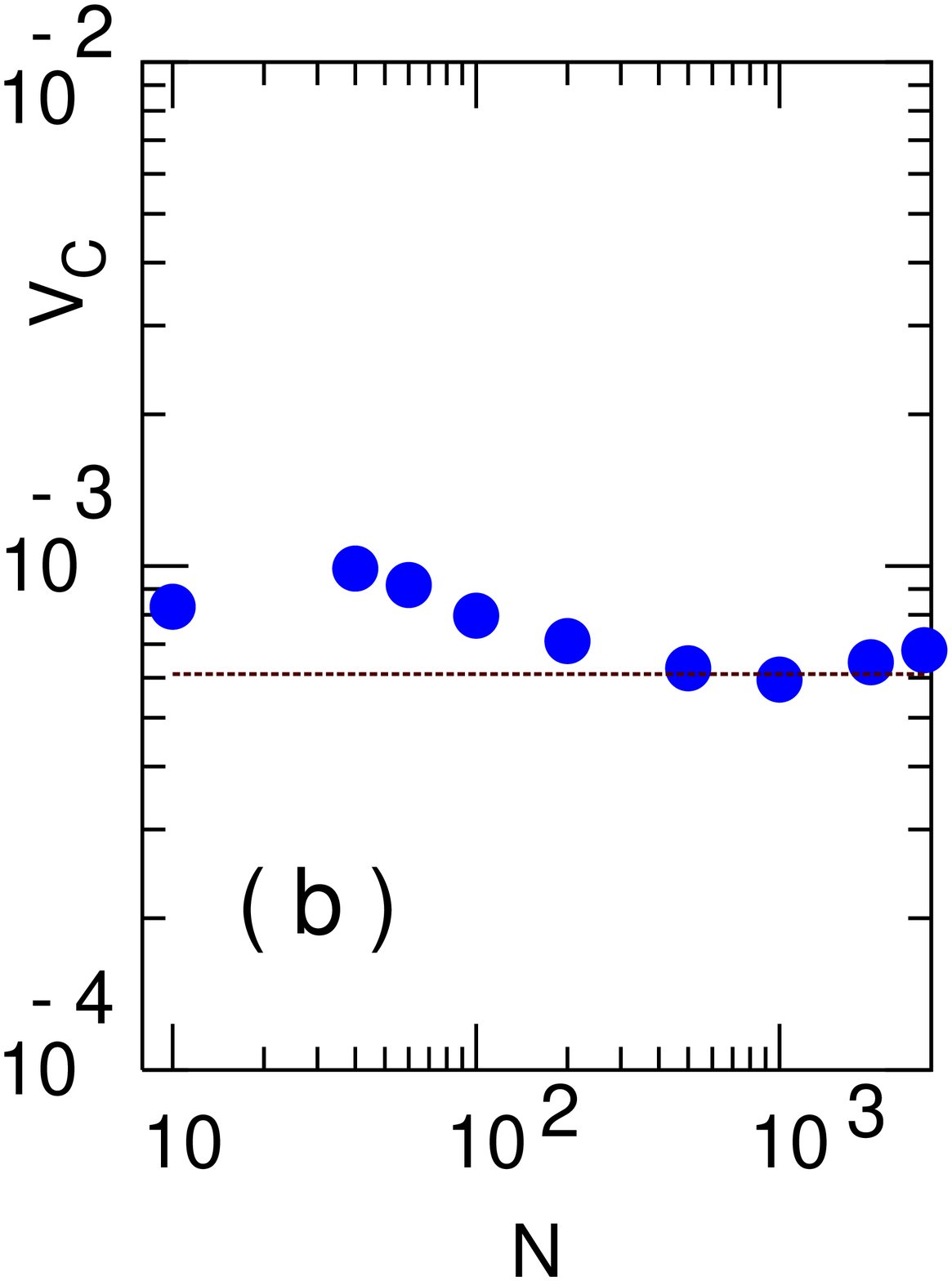}}
\end{tabular}
\caption{(a) $R_m$, $R_{\chi}$ versus $N$ at $T_C$ and (b) $V_C$
  versus $N$ at $T_C$.The straight lines show straight line fits to
  $R_m$ and $R_{\chi}$ . } 
\label{SELFAVG}
\end{center}
\end{figure}}

\newcommand{\SINGLE}{%
\begin{figure}
\begin{center}
\begin{tabular}{cc}
\resizebox{42mm}{!}{\includegraphics{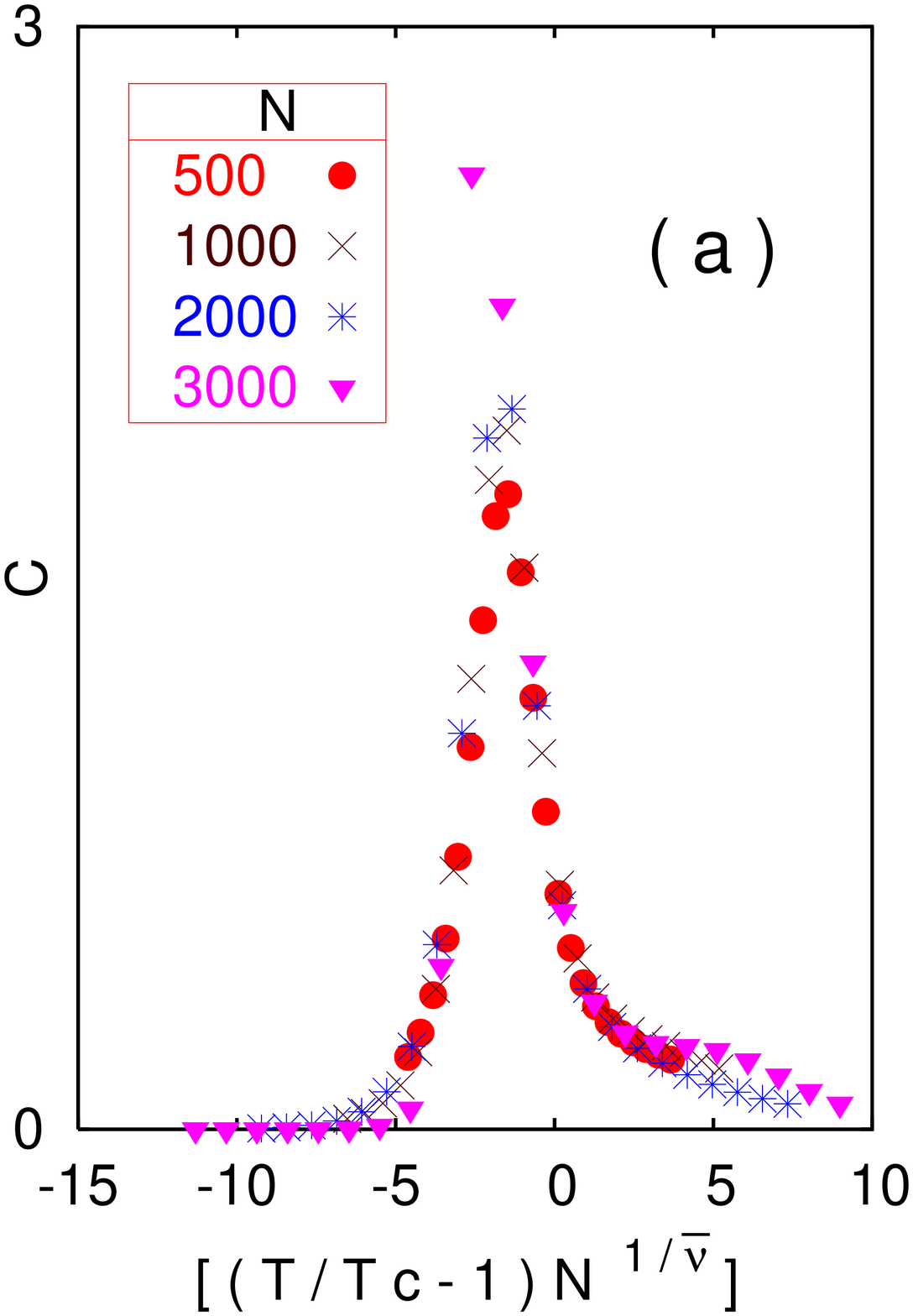}}&
\resizebox{42mm}{!}{\includegraphics{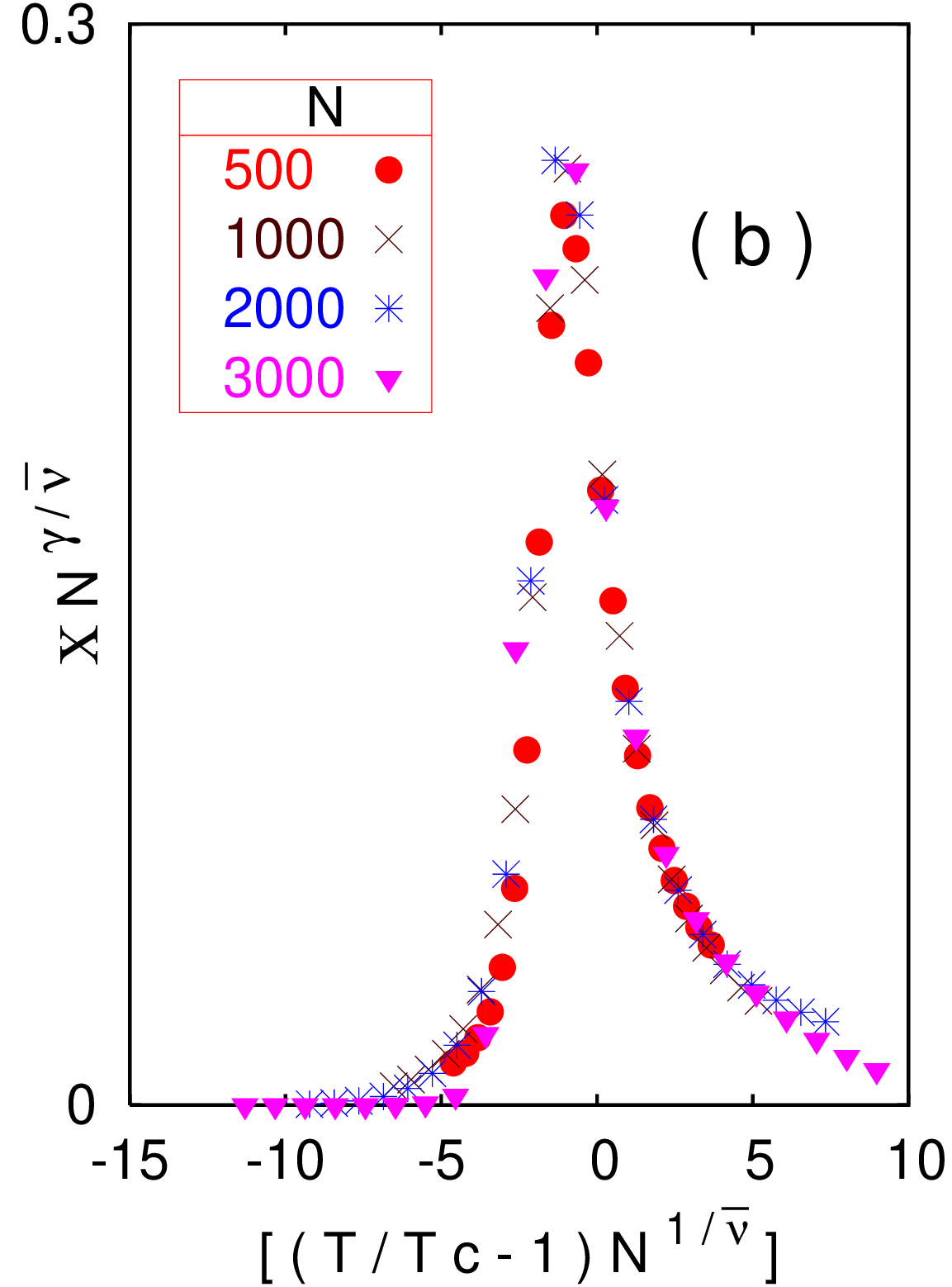}}\\
\end{tabular}
\caption{Data collapse of $C$ and $\chi$ on considering a single realisation of randomness}
\label{SINGLE}
\end{center}
\end{figure}}

\begin{abstract}
  Recent renormalization group results predict non self averaging
  behaviour at criticality for relevant disorder.  However, we find
  strong self averaging(SA) behaviour in the critical region of a
  quenched Ising model on an ensemble of small-world networks, despite
  the relevance of the random bonds at the pure critical point. 
\end{abstract}

\maketitle

\section{Self-averaging}

Very often in physics one encounters situations where quenched
disorder plays a prominent role. Any physical property $X$ of such a
disordered system, therefore, requires an averaging over all
realisations. It would suffice to have a description in terms of the
average $[X]_{\rm av}$ where $[. . ]_{\rm av}$ denotes averaging over
realisations (``sample averaging'') provided the relative variance
$R_X=V_X/[X]^2_{\rm av}\to 0$ for large $N$, where $V_X=[X^2]_{\rm
  av}-[X]^2_{\rm av}$.  In such a case a single large system is enough
to represent the ensemble. Such a quantity is called self-averaging.
Off criticality, if one builds up a large lattice from smaller blocks,
then thanks to the additivity property of an extensive quantity,
central limit theorem guarantees that $R_X\sim N^{-1}$ ensuring self-averaging.  
In contrast,  at a critical point, because of long range correlations the answer
to the question as to whether $X$ is self-averaging or not becomes non-trivial.

Randomness at a pure critical point is classified as relevant or
irrelevant if, following the standard definition of relevance, it
changes the critical behaviour (i.e. the critical exponents) of the
pure system.  Recent renormalization group and numerical studies have
shown that if randomness or disorder is relevant, then
{\it{self-averaging property is lost}}\cite{AH-PRL-96,WD-PRE-98}.  In
particular, $R_X$ at the critical point {\it{approaches a constant}}
as $N\to\infty$.  Such systems are called non self-averaging.  A serious
consequence of this is that unlike the self-averaging case, even if the 
critical point is known exactly, statistics in numerical simulations cannot
 be improved by going over to larger lattices (large $N$).

Let us recollect the definitions of various types of self-averaging with the help of the asymptotic size dependence of a quantity like $R_X$. If $R_X$ approaches a constant as $N\to\infty$, the system is {\it non-self-averaging}  while if $R_X$ decays to zero with size, it is {\it self-averaging}.  Self-averaging systems are further classified as strong and weak. If the decay is $R_X\sim N^{-1}$ as suggested by the central limit theorem, mentioned earlier, the system is  said to be  {\it  strongly self-averaging}.     There is yet another class of systems which shows a slower power law decay $R_X\sim N^{-z}$ with $0< z<1$. Such cases are known as weakly self-averaging. The exponent $z$ is determined by the known critical exponents of the system.

The prediction of non-self-averaging nature of critical quantities is
an extremely significant result coming from general renormalization
group arguments.  This basic result of Ref.  \cite{AH-PRL-96} and the
hypothesis of Ref.  \cite{WD-PRE-98} can be summarized as follows.
According to finite size scaling, when the critical region sets in, the
size of the system is comparable to the correlation length $\xi$ that
grows as the critical point is approached. The appropriate scaling
variable is $N/N_c$ where $N_c=\xi^d$ is the correlation volume in $d$
dimensions. At the critical point of a random system, there is an
additional source of fluctuation from the variation in the transition
temperature itself.  Therefore, instead of the conventional finite
size scaling, a sample dependent scaled variable is required.  A reduced
temperature is defined as $\tilde{t_i}=|T-T_c(i ,N)|/T_c $ where
$T_c(i ,N)$ is a pseudo-critical temperature of sample $i$ of $N$
sites with $T_c$ as the ensemble average of critical temperature in
the $N\rightarrow\infty$ limit.  In terms of this temperature, a
critical quantity $X$ is expected to show a sample dependent finite
size scaling form
\bege
\label{fss}
X_i(T ,N)=N^\rho Q (\tilde{t_i}N^{1/{\bar{\nu}}}) 
\ende 
where $\rho$ characterizes the behaviour of $[X]_{\rm av}$ 
at $T_c$.$^{\footnotemark[1]}$
{\footnotetext[1]{Conventional notations of critical exponents are
    used: $C \sim t^{-\alpha}$, $\chi \sim t^{-\gamma}$, $\xi \sim
    t^{-\nu}$ where $C$, $\chi$ and $\xi$ denote the specific heat,
    magnetic susceptibility and correlation length of the system. $t$
    is the temperature-like variable with the critical point at $t=0$.
  }}  Thus $\rho=\gamma/{\bar{\nu}}$ where ${\bar{\nu}}=d \nu$ when
$X$ is the magnetic susceptibility $\chi$.  The RG approach seems to
validate this hypothesis especially the absence of any extra anomalous
dimension in powers of $N$ for $R_X$.  Incidentally, this hypothesis,
Eq.  \ref{fss}, excludes rare events of large pure type lattices for
which pure $\bar\nu$ should be used.  We are not considering such
cases dominated by these rare events (Griffiths' singularity).  With
this scaling form, the relative variance $R_X$ at the critical point or
in the critical region is given by
\begin{equation}
\label{eq:2}
R_X \sim [(\delta T_c)^2]_{\rm av} N^{2/{\bar{\nu}}} ,
\end{equation}
where $[(\delta T_c)^2]_{\rm av}$ is the  variance of the
pseudo-critical temperature.  A random system can have several
temperature scales, namely $(T_c(N)-T_c)$ and $(T-T_c)$, in
addition to the shift in the transition temperature itself.  It is
plausible that for a system with relevant disorder all these scales
behave in the same way so that typical fluctuations in the
pseudo-critical temperature is set by the correlation volume,
yielding, 
\begin{equation}
  \label{eq:8}
[(\delta T_c)^2]_{\rm av} \sim N^{-2/{\bar{\nu}}}
\end{equation}
An immediate consequence of this is that $R_X$ of Eq. \ref{eq:2}
approaches a constant as $N\rightarrow\infty$ indicating complete
absence of self-averaging at the critical point in a random system.
These predictions have been verified for various types of relevant and
irrelevant disorders and also with canonical(ensemble of fixed
concentration of disorder) and grand canonical disorder 
at the random critical point
\cite{WD-PRE-98,Dillmann,Marques1,Marques2} .

\section{Small-world network}

Over the last few years, small-world networks (SWN) have emerged as a
new class of graphs with characteristic statistical properties defined
over the ensemble of networks.  Starting from an Euclidean lattice,
one may obtain an SWN by rewiring the original lattice or by the
addition of random long range bonds, even with a sufficiently small
probability $p$\cite{Strogatz,Albert,Kulkarni}.  The network is so
named because any two points far away on the underlying lattice can be
bridged, on the average, by a finite number of connections.  It has
been observed that for an Ising model defined on such an ensemble of
graphs, this set of random bonds changes the critical behaviour to the
mean field type in all the models studied so
far\cite{Barrat,Gitterman,Hastings,Herrero,Hong}. This, then, by
definition, makes this set of random bonds, added to the underlying
``pure'' lattice, a ``relevant'' variable.  There is albeit a debate
on the crossover exponent for $p$ at the pure critical point, in the
limit $p\to0$\cite{Hastings,Herrero}. Since quenched averaging is
important, replica trick has been resorted to \cite{Barrat,Gitterman}.
Overlaps and other quantities of interest for  general quenched random
systems have also been studied\cite{Nikoletopoulos}.

The aforementioned result of non-self-averaging would imply a strong
influence of the network i.e. a network to network variation of an
extensive quantity at the shifted critical point.  This aspect of
disorder of the small-world networks is the primary motivation of the
study reported here.

With this background we set to check the behaviour of $R_X$ for
various $X$ for SWN.  Some of the major differences with respect to
the previous studies of random systems may be mentioned here.  By
construction, the random bonds in SWN introduce long-range
interactions unlike the short range models studied so far.
{ \it The  self-averaging behaviour of  long range cases is of
importance \cite{Marques1,Marques2} because of the specialities  known,
e.g., for disorder with long range correlations\cite{Halperin}}.
We would also like to note that compared to many other random systems,
the exact values of the changed exponents are known in the SWN case.
It is well established that the shifted critical behaviour has 
$\alpha=0$ and therefore $\bar\nu = 2$.

\section{The Ising model on a Small-World network}

Let us think of the ferromagnetic Ising model with  spins $s_i=\pm 1$
at each site $i$ of a network based on an Euclidean lattice of $N$
points similar to Ref. 
\cite{Barrat,Gitterman,Hong,Herrero,Hastings,Lopez,Nikoletopoulos}
. Its Hamiltonian is taken as 
\begin{equation}
\label{eq:1}
H({\cal \{S\}}) = - J \sum_{\langle ij\rangle} s_i s_j - J \sum_{(ij)}s_is_j ,
\end{equation}
where $\langle ij\rangle$ are the nearest-neighbours on the lattice ,
$(ij)$ are the long distance neighbours along the random bonds of
$\cal \{S\}$ added for the network and $J>0$.  The Hamiltonian, being
dependent on the set ${\cal \{S\}}$, is random for a given
configuration of the spins $\{s_i\}$.  Any physical property $X$ of
the model, therefore, requires an overall averaging over the ensemble
of networks.  For our studies we need the shifted critical point and
the sample to sample fluctuations of various quantities at this
critical point.  Since the random bonds are known to be relevant,
for all $d\le 4$, there is no loss of generality if the SWN is built from a
one dimensional lattice.

We start with the Ising model on a SWN in 1D. In our model each site
on the lattice with an Ising spin has random links to two distant
spins such that no two spins are connected by more than one link. All
links of equal strength. Thus we have a ``canonical" scenario since
the number of links at each site is fixed. Hence no extra
normalisation factor is needed in the long range part of the
Hamiltonian of Eq {\ref{eq:1}}.  In the present work we chose
$J/k_{B}=1$ where $k_{B}$ is the Boltzmann constant.

\section{Simulation details}

\BINDER

Data were taken at $T=2.85$ (close to the estimated critical
temperature)and $\chi$, $C$ and the Binder
cumulant $U_{m}=[{<m^4>}/{<m^2>}^2]_{\rm av} -3$ were calculated, using the
single histogram reweighting technique \cite{Ferrenberg}. We examined
lattice sizes $N=100, 500, 1000, 2000, 3000$ in our Monte Carlo
simulations.  We studied $1535$ samples for $N=100$ to $517$ samples
for $N=3000$ using $10^3$ equilibration and $10^6$ MC steps for each
$N$. Data were taken at intervals of $10^3$ MC steps .

A data-collapse of $U_m$ with finite size scaling variable
$N^{1/\bar\nu}(T-T_c)/T_c$ would give $T_c$ (the infinite lattice
critical temperature) and $\bar\nu$.  By using the data-collapse
method of Ref. \cite{Seno} we obtained $T_c=2.83(2)$ and
$1/\bar\nu=0.50(1)$
  
The value of $\bar\nu$ is consistent with previous results
\cite{Hong}.  The resulting collapse is shown in Fig.
\ref{fig:BINDER}.  We also investigated similar plots for $\chi$ and
$C$ after averaging over many realisations of disorder (not shown
here), with same results.

Further proof of the mean-field nature of the transition comes from
the comparison of the data with the mean-field form of $U_m$. To
evaluate the mean field form of $U_m$ we use the mean-field form of
the magnetisation per spin m probability distribution in the critical
region \cite{Binder} 
\bege
\label{binder1}
P_N(m)\propto \exp { [ -N(a_1t_Nm^2+a_2m^4) ] } 
\ende 
with $t_N$ being the critical temperature of a lattice of 
size $N$ and $a_1$, $a_2$
being constants. By replacing $\hat{m}=(a_2N)^{1/4}m$, we find $U_m$
where the averages are obtained by integrating $\hat{m}$ from
$-\infty$ to $+\infty$ with the weight 
\bege
\label{binder2}
P_N(\hat{m})\propto \exp { [ -b_1(x-b_2)\hat{m}^2-\hat{m}^4 ] } 
\ende
where $b_1=a_1/{a_2}^{1/2}$ , $x$ is the finite size scaling variable
$[(T-T_c)/T_c]N^{1/{\bar\nu}}$ with $b_1b_2$ taking care of the finite
size shift of the critical temperature.  The solid curve in Fig.
\ref{fig:BINDER} is obtained with $b_1=1. 7$, $b_2=0. 72$.

We find good data collapse by using $\tilde t=T-T_c$ even though
finite size scaling is supposedly better with the use of $\tilde
t=T-T_c(N)$ after finding out the $T_c(N)$ for every sample
size\cite{Bernardet}. This is because our system, as we show, is
self-averaging and this method should be more pertinent for non- or
weakly self-averaging systems.
\DISTRIBUTION
To investigate the distribution of pseudo-critical temperature,
$\beta_c(i,N)$ (the temperature at which the specific heat of sample
$i$ of size $N$ is a maximum), data were taken at $T=2.85$ and
$\beta_{c}(i ,N)$ for various $N$ were calculated using the histogram
method \cite{Ferrenberg}. The distribution of $\beta_{c}(i ,N)$ for
$N=10,50,100$ is constructed.We studied $329$1 lattice samples for
$N=10$, $1645$ lattice samples for $N=50$ and $1535$ lattice samples
for $N=100$.  We find that the inverse critical temperature
$\beta_{c}(i ,N)$, scales as 
\bege
\label{sa-scaling} 
[(\beta_{c}(i ,N)-\beta_{c}(N))^2]_{\rm av} \sim N^{-2/{\bar{\nu}}} , 
\ende 
which is consistent with Eq. (\ref{eq:8}), 
but one also needs a scale factor $N^{-1/2}$ for the probability
distribution $P(\beta_c(i,N))$. Fig.  \ref{fig:DISTRIBUTION} shows the
data collapse of this distribution.  The data-collapse is best
achieved with $\bar\nu=2.00(3)$ which is consistent with the value of
$\bar\nu$ obtained in the data collapse shown in fig. \ref{fig:BINDER}.
\SELFAVG
We then studied $R_m$, $R_{\chi}$ and $V_C$ at $T_c^\infty$ for the
above lattice sizes. About 56440 lattice samples for N=10 to 1000
lattice samples for N=3000 were studied. For each sample we used
$10^3$ equilibration and $10^5$ MC steps. Data were taken at intervals
of $10^3$ MC steps. The data is fitted to the form $R_X=A_X \it
N^{\rho_{X}}$ where $R_X$ is the relative variance for $m$ and $\chi$.
The values obtained are $\rho_{m}=-0.96(9)$, and
$\rho_{\chi}=-0.94(8)$. Thus $\chi$ and $m$ are {\it{strongly \sa}}.
The singular part of energy cannot be filtered out and hence the
behaviour of $V_E$ can not be predicted decisively. We see in Fig. 3
that $V_C$ is a constant as expected and hence C is also {\it{strongly
    \sa}}.
\SINGLE
\section{Discussion}

The fact that the peak scales inversely as the width shows that
despite the fluctuation in pseudo-critical temperatures,the
distribution approaches a $\delta-$function. As a result the critical
temperature of a large $N$ network can be thought of as the average of
pseudo-critical temperatures of the small sub-networks.This averaging
out is tantamount to \sa .This is in marked contrast to the other
cases of random systems studied so far {\cite{WD-PRE-98,Dillmann,Marques1,Marques2}. It is tempting to conclude that in addition to relevant randomness, a
broad distribution of pseudo-critical temperature is a requirement for
non-self-averaging.

Whilst in the present work we have used a ``canonical" ensemble with a
fixed number of bonds,  small scale simulations of the Ising model 
on a SWN in a ``grand canonical" ensemble, where the number of bonds
can vary, also indicated self-averaging as found here.

In case of a strongly \sa system, a typical sample should be a
representative of the average.  We observe good data collapse with
even a single realisation of disorder (as shown in Fig. \ref{SINGLE}).
Thus, in the light of the present work, in such situations, an
annealed averaging as done in ref \cite{Gitterman} should work well.
Consequently no extra order parameter like approach should be needed
for networks .

It is not clear if this feature of strong \sa is a consequence of
$\alpha=0$, in which case it could be true for all relevant disorder
problems with mean-field behaviour.  An extension of the RG
argument\cite{AH-PRL-96}  to encompass situations with
sharp limit of $T_c$ distribution  and long range
interactions, may shed light on this.  Whether
this result on the disorder aspect of a network is important in other
real life situations like the railway network \cite{Sen} needs further
study.  

To conclude, we investigated the self-averaging behaviour of the Ising
model on a small world network.  The distribution of $\beta_c(i,N)$ is
found to become sharper as $N\to\infty$ with the fluctuation decaying
as $[(\delta \beta_c(i ,N))^2]_{\rm av}\sim N^{-2/\bar \nu}$.  The
data collapse of various physical quantities both for a single
realisation of disorder and after averaging over many disorder
realisations showed no significant difference.  At $T_c^\infty$, the
relative fluctuations $R_m$, $R_\chi$ for magnetization and
susceptibility are found to behave as $R_m, R_{\chi} \sim N^{-1}$
while the variance $V_C$ for the specific heat approaches a constant
for large $N$.  Hence the system is strongly self-averaging in the
critical region in spite of relevant randomness.  Our results have the
following implications. From the small-world networks perspective, the
random or statistical features of the network do not play a role in
the long range behaviour at the critical point so that one may replace
the ensemble of small-world networks by a single average network.

\end{document}